\documentclass[english,review]{elsarticle}
\usepackage{mathpazo}

\usepackage[T1]{fontenc}
\usepackage[latin9]{inputenc}
\usepackage{geometry}
\usepackage{color}
\usepackage{bm}
\usepackage{amsmath}
\usepackage{amssymb}
\usepackage{cancel}
\usepackage{graphicx}

\makeatletter

\usepackage{xcolor}

\newcommand{\rttensor}[1]{\overline{\overline{\mathbf{#1}}}}

\@ifundefined{showcaptionsetup}{}{%
 \PassOptionsToPackage{caption=false}{subfig}}
\usepackage{subfig}
\makeatother

\usepackage{babel}
\begin{document}

\title{An unsupervised machine-learning checkpoint-restart algorithm using
Gaussian mixtures for particle-in-cell simulations}

\author[]{G. Chen\corref{cor1}}

\ead{gchen@lanl.gov}

\author[]{L. Chac\'{o}n}

\author[]{T. B. Nguyen}

\cortext[cor1]{Corresponding author}

\address{Los Alamos National Laboratory, Los Alamos, NM 87545}
\begin{abstract}
We propose an unsupervised machine-learning checkpoint-restart (CR)
algorithm for particle-in-cell (PIC) algorithms using Gaussian mixtures
(GM). The algorithm features a particle compression stage and a particle
reconstruction stage, where a continuum particle distribution function
(PDF) is constructed and resampled, respectively. To guarantee fidelity
of the CR process, we ensure the exact preservation of invariants
such as charge, momentum, and energy for both compression and reconstruction
stages, everywhere on the mesh. We also ensure the preservation of
Gauss' law after particle reconstruction. As a result, the GM CR algorithm
is shown to provide a clean, conservative restart capability while
potentially affording orders of magnitude savings in input/output
requirements. We demonstrate the algorithm using a recently developed
exactly energy- and charge-conserving PIC algorithm using both electrostatic
and electromagnetic tests. The tests demonstrate not only a high-fidelity
CR capability, but also its potential for enhancing the fidelity of
the PIC solution for a given particle resolution.
\end{abstract}
\maketitle

\section{Introduction}

Resiliency, data locality, and asynchrony are key major challenges
facing the practical use of exascale computing for scientific applications.
Because of extreme concurrency, very large system scale, and complex
memory hierarchies, hardware failures (both ``soft'' and ``hard'')
are expected to become more frequent towards and beyond exascale.
Currently, 100 billion transistors/node , thousands of nodes, 10M-core
supercomputers are built (e.g. Summit and Sierra \citep{kahle2019}).
The very large total number of components will lead to frequent failures,
even though the mean time between failures (MTBF) for the individual
components may be large. For instance, while the MTBF of a CPU can
be months to years \citep{nightingale2011cycles}, that of current
supercomputers can be within a few hours \citep{liu2018large,rojas2019analyzing}.
With billion-core parallelism at exascale, the MTBF has been projected
to be within (or even far below) one hour \citep{dauwe2017analysis,miao2018energy}.
Therefore, it is important to enable efficient strategies that allow
software and algorithms to perform in a frequently interrupted environment.

Particle-based simulation algorithms are widely employed, at the heart
of many algorithmic strategies (e.g., Monte Carlo, particle-in-cell,
molecular dynamics) and applications (e.g., aerosol transport in combustion
and climate, radiation transport, and plasma transport). Checkpoint/restart
(CR) enables simulation recovery from previous interrupted simulations
due to either finite queue wall-clock-time limits or hardware (HW)
failures. This is commonly done by storing a sufficiently complete
data snapshot to disk at given time intervals, which can be then read
back to restart the simulation. Particle-based simulations at the
extreme scale are particularly challenged by the input/output (IO)
requirements of storing billions to trillions of particles, as is
already the case in the leading-class supercomputers. The challenges
are significantly worsened by the current trend towards hierarchical
architectures, featuring many levels of parallelism, each delivered
by different architectural solutions. Synchronous checkpointing in
hierarchical systems would require bulk synchronization across the
levels of the hierarchy, and ultimately storage in the file system
via IO. Asynchronous IO, as well as memory-based IO, are being explored
as partial solutions to the CR problem \citep{lofstead2008flexible}.
Nevertheless, \emph{any} IO-based CR strategy would greatly benefit
from a high-fidelity compression strategy of data for particle simulations.

In this study, we explore the viability of an \emph{unsupervised machine-learning,
optimization-based CR strategy} for plasma particle-in-cell (PIC)
simulations, combining \emph{optimal} (in some sense, to be clarified
below) compression and reconstruction of particle data. Compression
of particle data is performed per spatial cell by construction of
a continuum particle distribution function (PDF) with a Gaussian mixture
\citep{mclachlan2004finite}, based on a penalized maximum-likelihood-estimation
(PMLE) approach using complexity criteria \citep{figueiredo2000unsupervised}.
The resulting optimization problem is solved by an adaptive Expectation-Maximization
(EM) algorithm \citep{dempster1977maximum,figueiredo2000unsupervised},
which can automatically search for the optimal number of Gaussian
components satisfying a generalized ``minimum-message-length (MML)''
Bayesian Information Criterion \citep{wallace2005statistical}. The
method can be formulated to conserve up to second moments exactly
\citep{behboodian1970mixture}. Particle-data is reconstructed (also
locally per cell) by sampling of the PDF in velocity space (here using
Monte Carlo), with a simple moment-matching projection technique \citep{lemons2009small}.
Particle spatial positions within a given cell are re-initialized
randomly (i.e, we assume that the plasma is uniform within a cell).
Both compression and reconstruction operations are local in configuration
space (i.e., each computational cell features an \emph{independent}
PDF reconstruction process) and done \emph{in-situ} (assuming that
the cell has sufficient particles, e.g., more than 10), and only Gaussian
parameters are checkpointed.

A quiescent restart in plasmas requires, in addition to the preservation
of (at least) moments up to second order, the enforcement of Gauss'
law (i.e., $\nabla\cdot\mathbf{E}=\rho$ where $\mathbf{E}$ is electric
field, and $\rho$ is charge density) discretely everywhere on the
spatial mesh. Gauss' law is closely related to charge conservation,
and local violations will result in plasma waves being launched to
equilibrate charge. The electric field is saved at CR, and is thus
available at both compression and reconstruction stages. To enforce
Gauss' law discretely, it is sufficient to ensure resampled particles
\emph{exactly} match the charge density field per species and per
cell. We accomplish this by correcting particle weights according
to a straightforward mass-matrix solve \citep{burgess1992mass}.

The potential for IO compression of particle data using GM is quite
large. Each Gaussian component of the mixture requires ten parameters
to be fully specified, which is comparable to the number of degrees
of freedom needed per particle in a 3D-3V PIC method (e.g., three
positions and three velocities, plus particle weight and optionally
an integer identifying the cell on the mesh). Given that a few Gaussians
(< $10$) are usually sufficient (as demonstrated in our numerical
tests) to capture most details of the PDF, and that typical PIC simulations
employ hundreds if not thousands of particles per cell, it follows
that GM can easily result in several orders of magnitude savings in
IO requirements for checkpointing particle data.

The proposed CR strategy exactly conserves local (per cell) charge,
momentum, and energy, satisfies Gauss' law everywhere, is massively
parallel, communication-avoiding, locality-aware, and asynchronous
by construction (except for the mass-matrix solve step), and only
synchronizes and checkpoints compressed data. It is worth pointing
out that we are not the first ones to realize the potential of GM
PDF reconstruction in PIC algorithms, with various authors having
used it in the past for diagnostics \citep{dupuis2020characterizing},
to couple with other physical processes \citep{bowers2004maximum},
or, more related to this study, for Gaussian-to-Gaussian remapping
in 1D-1V phase-space to eliminate Gaussian-shape distortion in a finite-mass-method-based
Vlasov-Poisson algorithm \citep{larson2015finite}. However, to our
knowledge, this is the first application of an \emph{adaptive} GM
algorithm for particle data compression in CR of PIC simulations.

The rest of the paper is organized as follows. Section \ref{sec:Methods}
introduces the basic concepts of the PMLE method employed in this
study to learn and resample the Gaussian mixture, including the strategies
to enforce Gauss' law and conserve of up to second moments. Section
\ref{sec:results} demonstrates the CR algorithm for prototypical
plasma-physics electrostatic and electromagnetic PIC tests, and demonstrate
the potential of GM to improve the PIC solution for a given particle
resolution. We also explore ways to improve the efficiency of the
underlying EM algorithm to find the GM. Finally, we conclude in Section
\ref{sec:conclusions}.

\section{Methodology}

\label{sec:Methods}

We describe the two main elements of the CR GM strategy, namely, GM
component estimation (particle-data compression) and GM sampling (particle-data
reconstruction). Specifically, an adaptive EM algorithm is used to
estimate the number of components of the Gaussian mixture and their
parameters, and a moment-matching sampling technique is used to regenerate
particles from the Gaussian mixture.

A GM is defined as a convex combination of $K$ Gaussian distributions:
\begin{equation}
f(\mathbf{x})=\sum_{k=1}^{K}\omega_{k}f_{k}(\mathbf{x}),\label{eq:mix-dist}
\end{equation}
where each Gaussian $f_{k}$ is weighted by $\omega_{k}$ with $\sum_{k}\omega_{k}=1$
and $w_{k}>0$. The Gaussian distribution is defined as 
\begin{equation}
f_{k}(\mathbf{x})=\frac{1}{\sqrt{(2\pi)^{D}|\rttensor{\mathbf{\boldsymbol{\Sigma}}}_{k}|}}e^{-(\mathbf{x}-\boldsymbol{\mu}_{k})^{T}\rttensor{\mathbf{\boldsymbol{\Sigma}}}_{k}^{-1}(\mathbf{x}-\boldsymbol{\mu}_{k})/2},
\end{equation}
where $\boldsymbol{\mu}$ is a $D$-dimensional mean vector, $\rttensor{\mathbf{\boldsymbol{\Sigma}}}$
is a $D\times D$ covariance matrix, and $|\rttensor{\mathbf{\boldsymbol{\Sigma}}}|$
is the determinant of $\rttensor{\mathbf{\boldsymbol{\Sigma}}}$.

\subsection{Adaptive GM models and the penalized maximum likelihood function}

The goal is to estimate the parameters $\bm{\theta}\equiv\{\boldsymbol{\omega},\boldsymbol{\mu},\overleftrightarrow{\boldsymbol{\Sigma}}\}$
of each Gaussian as well as the number of mixture components, $K$,
given $N$ independent samples $\mathbf{X}$$=(\mathbf{x}_{1}...\mathbf{x}_{N}$)
drawn from $f(\mathbf{x})$. Conventionally, maximum likelihood is
used to estimate $\bm{\theta}$ for a prescribed number of components
\citep{everitt2014finite}. However, estimating the number of components
itself is in fact also important, and can be addressed in the framework
of the Bayesian Information Criterion \citep{mclachlan2004finite}.
In what follows, we give a brief overview of this approach.

We seek to find the maximum likelihood of the model $K$ (given a
data set $\mathbf{X}$), which by Bayes' rule reads: 
\begin{equation}
p(K|\mathbf{X})=\frac{p(\mathbf{X}|K)p(K)}{p(\mathbf{X})},\label{eq:model posterior}
\end{equation}
where $p(K)$ is the prior probability distribution for the model
family, and $p(\mathbf{X})=\sum p(\mathbf{X}|K)p(K)$ is a normalizing
constant. If we assume that all models are equally likely \textit{a
priori}, then $p(K)$ is uniform. Therefore, maximizing $p(K|\mathbf{X})$
is equivalent to maximizing $p(\mathbf{X}|K)$, which is the so-called
marginal likelihood (also known as evidence \citep{mackay2003information}
or type II maximum likelihood \citep{good1965estimation}), and is
given by:
\begin{equation}
p(\mathbf{X}|K)=\int p(\mathbf{X}|\bm{\theta},K)p(\bm{\theta}|K)d\bm{\theta},\label{eq:marginal likelihood}
\end{equation}
where $p(\bm{\theta}|K)$ is a prior probability distribution, and
$p(\mathbf{X}|\bm{\theta},K)$ is the likelihood function, which for
a Gaussian mixture reads:
\[
p(\mathbf{X}|\bm{\theta},K)=\sum_{k=1}^{K}\omega_{k}f_{k}(\mathbf{x}_{i}|\boldsymbol{\mu}_{k},\rttensor{\mathbf{\boldsymbol{\Sigma}}}_{k}).
\]
We seek to maximize Eq. \ref{eq:marginal likelihood}. For completeness,
the derivation is carried out in \ref{app:PLF}, and results in the
penalized log-likelihood function:
\begin{equation}
L(\bm{\theta})=\mathrm{ln}\left[p(\mathbf{X}|\bm{\theta},K)\right]-\frac{d}{2}\mathrm{ln}N-\frac{T}{2}\sum_{i=1}^{K}\mathrm{ln}(\omega_{i}),\label{eq:Simple penalized likelihood}
\end{equation}
The last term of Eq. \ref{eq:Simple penalized likelihood} is crucial
for finding the number of components, and avoiding over-fitting and
singularities of standard maximum likelihood estimate (MLE) \citep{figueiredo2000unsupervised}.
As pointed out in Ref. \citep{figueiredo2000unsupervised}, this term
is an effective Dirichlet prior with negative parameters. Such a prior
has a strong tendency to annihilate redundant components. We refer
to Ref. \citep{rousseau2011asymptotic} for a theoretical treatment
on this important point, and Refs. \citep{zivkovic2004improved,tu2016modified}
for its practical use in the context of Gaussian mixtures. 

\subsection{Learning the GM model by a penalized MLE}

We follow the standard method of MLE to seek optimum values of the
Gaussian parameters. This is achieved by maximizing the penalized
likelihood function, Eq. \ref{eq:Simple penalized likelihood}. For
a given set of particles, the penalized log-likelihood function is
given by 
\begin{equation}
L(\bm{\theta})=\sum_{p=1}^{N}\alpha_{p}\mathrm{ln}\left[\sum_{k=1}^{K}\omega_{k}f_{k}(\mathbf{v}_{p}|\boldsymbol{\mu}_{k},\rttensor{\mathbf{\boldsymbol{\Sigma}}}_{k})\right]-\frac{d}{2}\mathrm{ln}N-\frac{T}{2}\sum_{k=1}^{K}\mathrm{ln}(\omega_{k}),\label{eq:lnL-GM}
\end{equation}
where $\mathbf{v}_{p}$ is particle velocity and $\alpha_{p}$ is
the particle weight, which accounts for cases with non-identical samples
\citep{hasselblad1966estimation}. Note that $\sum_{p=1}^{N}\alpha_{p}=N$.
Typically, the MLE estimator is found by solving the likelihood equation:
\begin{equation}
\frac{\partial\mathrm{ln}L(\bm{\theta})}{\partial\bm{\theta}}=0,
\end{equation}
subject to the constraint that $\sum_{k}\omega_{k}=1$, with: 
\begin{equation}
\frac{\partial^{2}\mathrm{ln}L(\bm{\theta})}{\partial\bm{\theta}^{2}}<0.
\end{equation}

Setting the derivative of Eq. \ref{eq:lnL-GM} with respect to the
mean $\boldsymbol{\mu}_{k}$ of the Gaussian components to zero, we
obtain
\begin{equation}
\boldsymbol{\mu}_{k}=\frac{1}{N_{k}}\sum_{p=1}^{N}\gamma_{pk}\mathbf{v}_{p},\label{eq:muk}
\end{equation}
where 
\begin{equation}
\gamma_{pk}\equiv\frac{\alpha_{p}\omega_{k}f_{k}(\mathbf{v}_{p}|\boldsymbol{\mu}_{k},\rttensor{\mathbf{\boldsymbol{\Sigma}}}_{k})}{\sum_{k=1}^{K}\omega_{k}f_{k}(\mathbf{v}_{p}|\boldsymbol{\mu}_{k},\rttensor{\mathbf{\boldsymbol{\Sigma}}}_{k})},\label{eq:responsibility}
\end{equation}
and 
\[
N_{k}=\sum_{p=1}^{N}\gamma_{pk}.
\]
Note that 
\begin{equation}
\sum_{k=1}^{K}\gamma_{pk}=\alpha_{p}.\label{eq:Sum_responsibility}
\end{equation}
Setting the derivative of Eq. \ref{eq:lnL-GM} with respect to $\boldsymbol{\Sigma}_{k}$
to zero, we obtain 
\begin{equation}
\rttensor{\mathbf{\boldsymbol{\Sigma}}}_{k}=\frac{1}{N_{k}}\sum_{p=1}^{N}\gamma_{pk}(\mathbf{v}_{p}-\boldsymbol{\mu}_{k})(\mathbf{v}_{p}-\boldsymbol{\mu}_{k})^{\mathrm{T}}.\label{eq:sigmak}
\end{equation}
Maximizing Eq. \ref{eq:lnL-GM} with respect to the mixing coefficients
(again, constrained by $\sum_{k}\omega_{k}=1$) gives \citep{gauvain1994maximum,figueiredo2000unsupervised}:
\begin{equation}
\omega_{k}=\frac{N_{k}-\frac{T}{2}}{N-\frac{T}{2}K},\label{eq:omegak}
\end{equation}
provided $N_{k}-\frac{T}{2}>0$. This suggests one should begin with
more components than the ``true'' number of components of the mixture
\citep{rousseau2011asymptotic}. A component is eliminated ($K^{new}\leftarrow K^{old}-1)$
if $N_{k}-\frac{T}{2}\leq0$. In the limit of $N\rightarrow\infty$,
Eq. \ref{eq:omegak} recovers the standard MLE result, i.e., 
\begin{equation}
\tilde{\omega}_{k}=\frac{N_{k}}{N}.\label{eq:omegak-mle}
\end{equation}
The solution to Eqs. \ref{eq:muk}-\ref{eq:omegak} can only be found
iteratively.

\subsection{GM component estimation: Expectation-Maximization algorithm (EM-GM)}

\label{sec:The-EM-algorithm}

The EM-GM algorithm provides an iterative procedure to find a local
maximum of the log-likelihood function with respect to the Gaussian
parameters. For an extensive review of theoretical and practical aspects
of EM algorithm for finite mixture models, see Ref \citep{redner1984mixture}.
As discussed above, we start with a relatively large number of Gaussians
(\textasciitilde 10), for each Gaussian, we set its mean to coincide
with a randomly chosen particle, and its variance to be the same as
the total variance. Each EM-GM iteration consists of the following
steps \citep{celeux2001component}:

1. For each Gaussian component $k$, perform E-step: Given the parameter
set $\boldsymbol{\theta}_{k}^{it}$, where the superscript $it$ denotes
the iteration level, evaluate Eq. \ref{eq:responsibility}.

2. For the same Gaussian component $k$, perform M-step: Compute $\boldsymbol{\theta}_{k}^{it+1}=\{\boldsymbol{\omega},\boldsymbol{\mu},\rttensor{\mathbf{\boldsymbol{\Sigma}}}\}_{k}^{it+1}$
via Eqs \ref{eq:muk},\ref{eq:sigmak}, and \ref{eq:omegak}. If $\omega_{k}\leq0$,
remove the Gaussian, and let $K^{it+1}=K^{it}-1$, otherwise, $K^{it+1}=K^{it}$.

3. Re-normalize weight by $\omega_{k}=\omega_{k}/\sum_{i=1}^{K^{it+1}}\omega_{i}.$

4. Repeat steps 1 to 3 until all Gaussians are updated, check for
convergence by monitoring the log-likelihood function, Eq. \ref{eq:lnL-GM}.

Depending on how much the overlap of the Gaussians, convergence of
the algorithm may be slow. In general, convergence is slow when Gaussians
are poorly separated, and one should consider ways to accelerate it
for practical applications (see results in Sec. \ref{sec:results}).

\subsubsection{Properties of the EM-GM algorithm}

An important property of the EM-GM algorithm based on the \emph{unpenalized}
MLE is that it conserves up to second moments of the sample particles
\emph{exactly}, i.e., the mass, mean, and variance of the mixture
coincide with those of sample particles (see Ref. \citep{behboodian1970mixture}
for the 1D MLE, and derivation below for the multivariate case). As
a consequence, physical quantities such as mass, momentum, and energy
(or more precisely the pressure tensor) are conserved by the GM continuum
reconstruction of the particle PDF. However, such conservation property
is not inherited by the EM-GM algorithm based on the \emph{penalized}
MLE (for component adaptivity) . To recover the moment conservation
property, which is desirable for high-fidelity CR in particle simulations,
we perform the estimate in two steps: first use the PMLE to select
the optimal number of Gaussians, and then postprocess the result with
one step of unpenalized MLE to regain conservation.

It is useful to derive the conservation properties of the unpenalized-MLE-based
EM-GM algorithm as follows. We begin with the conservation of the
first moment (mean):
\begin{align}
E(\mathbf{v}) & =\sum_{k=1}^{K}\omega_{k}\boldsymbol{\mu}_{k}\nonumber \\
 & =\sum_{k=1}^{K}\omega_{k}\frac{1}{N_{k}}\sum_{p=1}^{N}\gamma_{pk}\mathbf{v}_{p}\nonumber \\
 & =\frac{1}{N}\sum_{p=1}^{N}\alpha_{p}\mathbf{v}_{p}=E(\mathbf{v}_{p}),\label{eq:equal-mean}
\end{align}
where we use the law of total expectation \citep{blitzstein2019introduction}
for the first equality, Eq. \ref{eq:muk} for the second equality,
and Eq. \ref{eq:omegak-mle} and \ref{eq:Sum_responsibility} for
the third equality. It is easily seen that, if Eq. \ref{eq:omegak}
is used instead of \ref{eq:omegak-mle}, the third equality above
would not follow through, breaking conservation of the first moment.

The derivation of the preservation of the second moments (variance)
follows a similar procedure,
\begin{align*}
Var(\mathbf{v}) & =E(Var(\mathbf{v}|\mathbf{y}))+Var(E(\mathbf{v|y}))\\
 & =\sum_{k=1}^{K}\omega_{k}\Sigma_{k}+E(E(\mathbf{v|y})^{2})-E(E(\mathbf{v|y}))^{2}\\
 & =\sum_{k=1}^{K}\omega_{k}\frac{1}{N_{k}}\sum_{p=1}^{N}\gamma_{pk}(\mathbf{v}_{p}-\boldsymbol{\mu}_{k})(\mathbf{v}_{p}-\boldsymbol{\mu}_{k})^{\mathrm{T}}+\sum_{k=1}^{K}\omega_{k}\mu_{k}^{2}-E(x)^{2}\\
 & =\frac{1}{N}\sum_{p=1}^{N}\alpha_{p}v_{p}^{2}-\left(\frac{1}{N}\sum_{p=1}^{N}\alpha_{p}\mathbf{v}_{p}\right)^{2}=Var(\mathbf{v}_{p}),
\end{align*}
where $\mathbf{y}$ is the hidden variable indicating the Gaussian
component that a particle belongs to. Here, the first equality is
the law of total variance \citep{blitzstein2019introduction}, the
second equality uses definitions of expectations and variances, and
the third equality uses Eq. \ref{eq:sigmak} and the so-called Adam's
law {[}i.e., $E(\mathbf{v})=E(E(\mathbf{v}|\mathbf{y})${]} \citep{blitzstein2019introduction}.
To get to the fourth equality, Eqs. \ref{eq:omegak-mle}, \ref{eq:Sum_responsibility},
and \ref{eq:equal-mean} are used. We observe that using Eq. \ref{eq:omegak}
instead of Eq. \ref{eq:omegak-mle} would again break the equality
of the variance between the Gaussians and particles. 

The derivations above indicate that Eq. \ref{eq:omegak-mle} is critical
for the conservation properties we wish to preserve for the Gaussians.
Equation \ref{eq:omegak} is, however, critical for selecting the
correct number of Gaussian components. We have designed a procedure
that combines the advantages of both (i.e., conservation \emph{and}
adaptivity) as follows. We first perform iterations using Eq. \ref{eq:omegak}
to prune out unnecessary Gaussians. Once converged, we simply perform
an extra step using Eq. \ref{eq:omegak-mle} for the mixing coefficients.
This is equivalent to accounting for the Gaussian weights based only
on the data, without penalization. Once we have Eq. \ref{eq:omegak-mle}
satisfied, the conservation properties are recovered as for the unpenalized
MLE case.

\subsection{GM component particle sampling}

\label{sec:sampling}

In physical space, we employ uniform random sampling independently
within each cell, which effectively assumes the PDF is constant within
each spatial cell. In velocity space (per spatial cell), we employ
the ancestral (or forward) sampling technique \citep{bishop2006pattern}
to generate random samples of a Gaussian mixture. This has the advantage
that it allows independent sampling per Gaussian while keeping sampled
particle weights identical. To begin, we re-write Eq. \ref{eq:mix-dist}
as
\begin{equation}
f(\mathbf{v})=\sum_{\mathbf{z}}f(\mathbf{z})f(\mathbf{v}|\mathbf{z}),\label{eq:mix-dist-1}
\end{equation}
where $\mathbf{z}$ is random unit vector of length $K$ (representing
the mixture components), with only one non-zero element $z_{k}=1$
(chosen randomly) \citep{bishop2006pattern}. The identification variable
$\mathbf{z}$ has a categorical distribution $f(\mathbf{z})$, and
the conditional distribution of $\mathbf{v}$ given $\mathbf{z}$
is a Gaussian. We first draw a sample from $f(\mathbf{z})$, which
identifies a Gaussian component $k$ with the probability $\omega_{k}$.
We then draw a sample from the multivariate Gaussian component \citep{tong2012multivariate}.
In this study, we have used the SPRNG scalable parallel library \citep{mascagni2000algorithm}
for random number generation.

Sampling errors in physical space result in violations of Gauss' law
(because the accumulated charge density on the mesh will not be identical
to the pre-checkpoint state). Similarly, sampling errors in velocity
space will break the conservation of mean and variance. Corrections
must be made to the sampling procedure to ensure that Gauss' law,
momentum and energy are exactly preserved \citep{lemons2009small}.
We discuss these next.

\subsubsection{Preservation of Gauss' law}

In plasmas, Gauss' law is directly related to local charge conservation.
After particles are spatially resampled within a cell, the local charge
density on the mesh no longer agrees with the pre-checkpoint stage.
The local charge density at cell $i$ is given by \citep{birdsall2004plasma}:
\[
\rho_{i}=\frac{1}{\Delta\mathbf{x}_{i}}\sum_{p}q_{p}\alpha_{p}S(\mathbf{x}_{i}-\mathbf{x}_{p}),
\]
with $q_{p}$ the particle charge, $\alpha_{p}$ the particle weight,
$\mathbf{x}_{p}$ the particle position within a cell, $\mathbf{x}_{i}$
the cell center, $\Delta\mathbf{x}_{i}$ the cell volume, and $S(\mathbf{x})$
a partition-of-unity interpolation kernel (typically a B-spline \citep{birdsall2004plasma}).
Clearly, changes in $\mathbf{x}_{i}$ will generally result in changes
in $\rho_{i}$, and therefore in Gauss' law, $\nabla\cdot\mathbf{E}=\rho$.

In order to recover the original charge density, we use a technique
introduced in Ref. \citep{burgess1992mass} to match the charge density
that before checkpointing. The basic idea here is to solve for a slight
adjustment of the weight of particles as follows. To be practical,
such a weight adjustment is assumed to be uniform within a cell. We
begin by assigning a weigh-correction degree of freedom per spatial
cell, $\delta A_{j}$, and define the particle weight correction for
all particles in cell $j$ to be equal to $\delta A_{j}$, i.e.:
\begin{equation}
\delta\alpha_{p}=\sum_{j}\delta A_{j}S_{0}(\mathbf{x}_{j}-\mathbf{x}_{p}),\label{eq:p-weight-correction}
\end{equation}
where $S_{0}(\mathbf{x}_{j}-\mathbf{x}_{p})$ is the zeroth-order
B-spline (top-hat) interpolation kernel. The new particle weight is
found as:
\begin{equation}
\alpha_{p}'=\alpha_{p}+\delta\alpha_{p}.\label{eq:new-p-weight}
\end{equation}
The weight correction $\delta\alpha_{p}$ is found by matching the
desired charge density $\rho_{i}'$ (here, using second-order B-splines),
i.e.:
\[
\rho_{i}'=\frac{1}{\Delta\mathbf{x}_{i}}\sum_{p}q_{p}\alpha_{p}'S_{2}(\mathbf{x}_{i}-\mathbf{x}_{p}).
\]
Introducing Eqs. \ref{eq:p-weight-correction} and \ref{eq:new-p-weight}
into the last equation, there results:
\[
\sum_{j}\delta A_{j}\underbrace{\sum_{p}q_{p}S_{0}(\mathbf{x}_{j}-\mathbf{x}_{p})S_{2}(\mathbf{x}_{i}-\mathbf{x}_{p})}_{M_{ij}}=\Delta\mathbf{x}_{i}(\rho_{i}'-\rho_{i}).
\]
The resulting linear system for $\delta A_{j}$ is a mass-matrix solve,
where the matrix is found from contributions from particles to each
cell according to stated interpolation rules. By construction, the
matrix $\rttensor{M}$ is sparse, diagonally dominant (because particles
in cell $j$ will contribute to that cell the most), and with all
positive entries. It is not stiff, and typically the resulting linear
system can be converged to round-off in a few iterations. 

\subsubsection{Preservation of mean and variance by local projection}

As a result of the mass-matrix Gauss correction step, changes in the
particle weight lead to changes in local momentum and energy, breaking
strict momentum and energy conservation in each cell. To recover
them, we use a well-known projection strategy of particle velocities
within each cell proposed by Lemons \citep{lemons2009small} to correct
for momentum and energy errors. 

To begin, we follow Ref. \citep{lemons2009small} and introduce a
scaling $\alpha$ and shift $\bm{\beta}$ for the particle velocity
as:
\[
\mathbf{v}_{p}^{\prime}=\alpha(\mathbf{v}_{p}+\bm{\beta}).
\]
The parameters $\alpha$ and $\boldsymbol{\beta}$ are determined
from the conservation constraints of momentum $\mathbf{p}$ and energy
$E$ (which are obtained from the GM):
\[
\sum_{p}\alpha_{p}\mathbf{v}_{p}=\sum_{p}\alpha_{p}'\mathbf{v}_{p}'=\mathbf{p}\,\,;\,\,\frac{1}{2}\sum_{p}\alpha_{p}v_{p}^{2}=\frac{1}{2}\sum_{p}\alpha_{p}'(v_{p}')^{2}=E.
\]
An exact solution $\alpha$ and $\boldsymbol{\beta}$ in terms of
$E$ and $\mathbf{p}$ can be found as:
\begin{equation}
\alpha=\sqrt{\frac{2EN_{p}'-p^{2}}{2E'N_{p}'-(p')^{2}}}\,\,;\,\,\boldsymbol{\beta}=\frac{\mathbf{p}-\alpha\mathbf{p}'}{\alpha N_{p}'},\label{eq:alpha}
\end{equation}
where:
\begin{equation}
N_{p}'=\sum_{p}\alpha_{p}'\,\,;\,\,\mathbf{p}'=\sum_{p}\alpha_{p}'\mathbf{v}_{p}\,\,;\,\,E'=\frac{1}{2}\sum_{p}\alpha_{p}'v_{p}^{2}.\label{eq:beta}
\end{equation}
Note that even though Schwarz inequality guarantees that $2E'N_{p}'\geq(p')^{2}$
(and therefore the denominator in Eq. \ref{eq:alpha} is always positive
definite), the numerator may occasionally become negative (we have
seen this when the number of particles is not large enough), and therefore
this is a potential failure mode of the approach. When this occurs,
there are two options: increase the targeted number of particles for
that cell, or forgo the local moment-matching step in that cell.

\section{Numerical experiments}

\label{sec:results}

In this section, we test the proposed CR algorithm using some prototypical
test problems, the 1D-1V electrostatic two-stream instability, and
the 2D-3V electromagnetic Weibel instability. We perform the simulations
with the DPIC code, based on a recently proposed implicit, charge
and energy conserving multidimensional electromagnetic PIC algorithm
\citep{chen2015multi}. Because of its exact charge- and energy-conserving
formulation, DPIC simulations represent a stringent test of the conservation
properties (or lack thereof) of the proposed CR algorithm.

\subsection{1D-1V two-stream electrostatic instability}

\begin{figure}
\begin{centering}
\includegraphics[width=0.8\columnwidth]{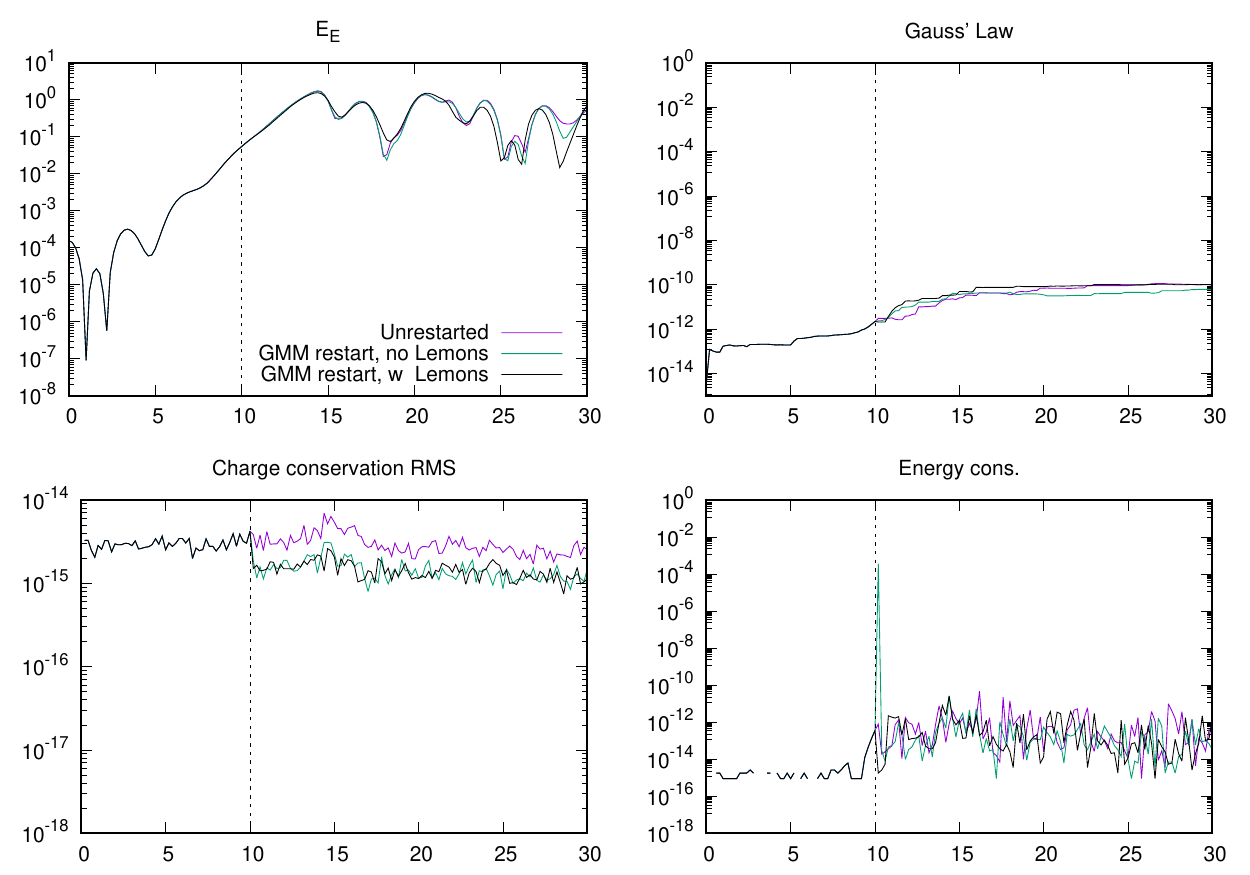}
\par\end{centering}
\caption{\label{fig:Two-stream-instability-restarted}Two-stream instability:
Semi-log-scale time history of the electric field energy $E_{E}$
(top-left), the rms of Gauss' law residual over the whole mesh (top-right),
the rms of the residual of the charge conservation equation (bottom-left),
and the change of total energy between subsequent time steps (bottom-right).
The simulations are obtained without restart, and with GM restart
at $t=10$ (in normalized units) with and without Lemons moment matching. }
\end{figure}

The two-stream instability \citep{lampert1956plasma} is an electrostatic
instability in which two counter-streaming particle beams exchange
kinetic and electrostatic energy, and as a result tangle up in to
a vortex in phase space \citep{roberts1967nonlinear}. The simulation
is performed for $L=2\pi$ (domain size, in Debye length units), $v_{b}=\sqrt{3}/2$
(beam speed, in electron thermal speed units), $N_{x}=32$ (number
of cells), $N_{p}=156$ (number of particles per cell), $\Delta t=0.2$
(time step in inverse plasma frequency units), with periodic boundary
conditions. Figure \ref{fig:Two-stream-instability-restarted} shows
the root-mean-square (rms) of the charge conservation equation residual
$(\partial_{t}\rho+\nabla\cdot\mathbf{j})$ over the mesh, the electric-field
energy $E_{E}=\sum_{i}\frac{E_{i}^{2}}{2}$, the total energy error
between subsequent timesteps, $|\mathcal{E}^{n+1}-\mathcal{E}^{n}|$,
with $\mathcal{E}$ the total sum of particle and electric-field energy,
and the rms of the residual of Gauss' law, $\nabla\cdot E-\rho$.
The plot compares the unrestarted run with two GM-restarted ones at
$t=10$ (mid/late linear stage), with and without Lemons' moment matching.
The results show exact conservation of charge for all cases, also
for energy except for the case where Lemons matching was not used
(which results in a large energy conservation error right after restart),
and excellent preservation of Gauss' law (commensurate with the nonlinear
tolerance). They also show excellent agreement in the temporal evolution
of the electrostatic field energy for all cases. For this run, the
GM algorithm is started with 8 Gaussian's per cell, resulting in an
average number of Gaussians per cell of 2, and therefore to an average
compression ratio of about 75. 

\begin{figure}
\begin{centering}
\includegraphics[angle=-90,width=0.4\columnwidth,trim={ 0.5in 0 0 0 },clip]{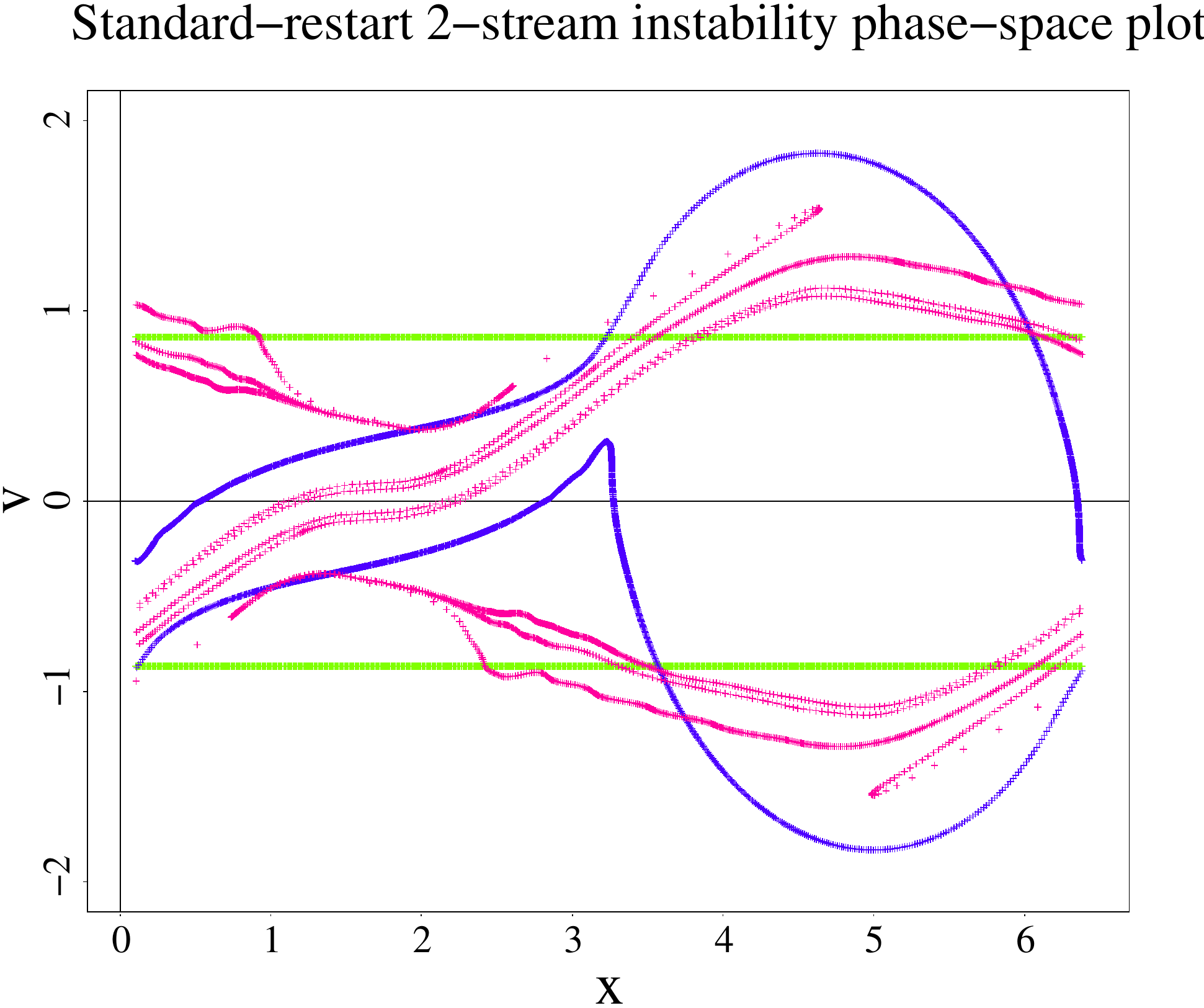}~\includegraphics[angle=-90,width=0.4\columnwidth,trim={ 0.5in 0 0 0 },clip]{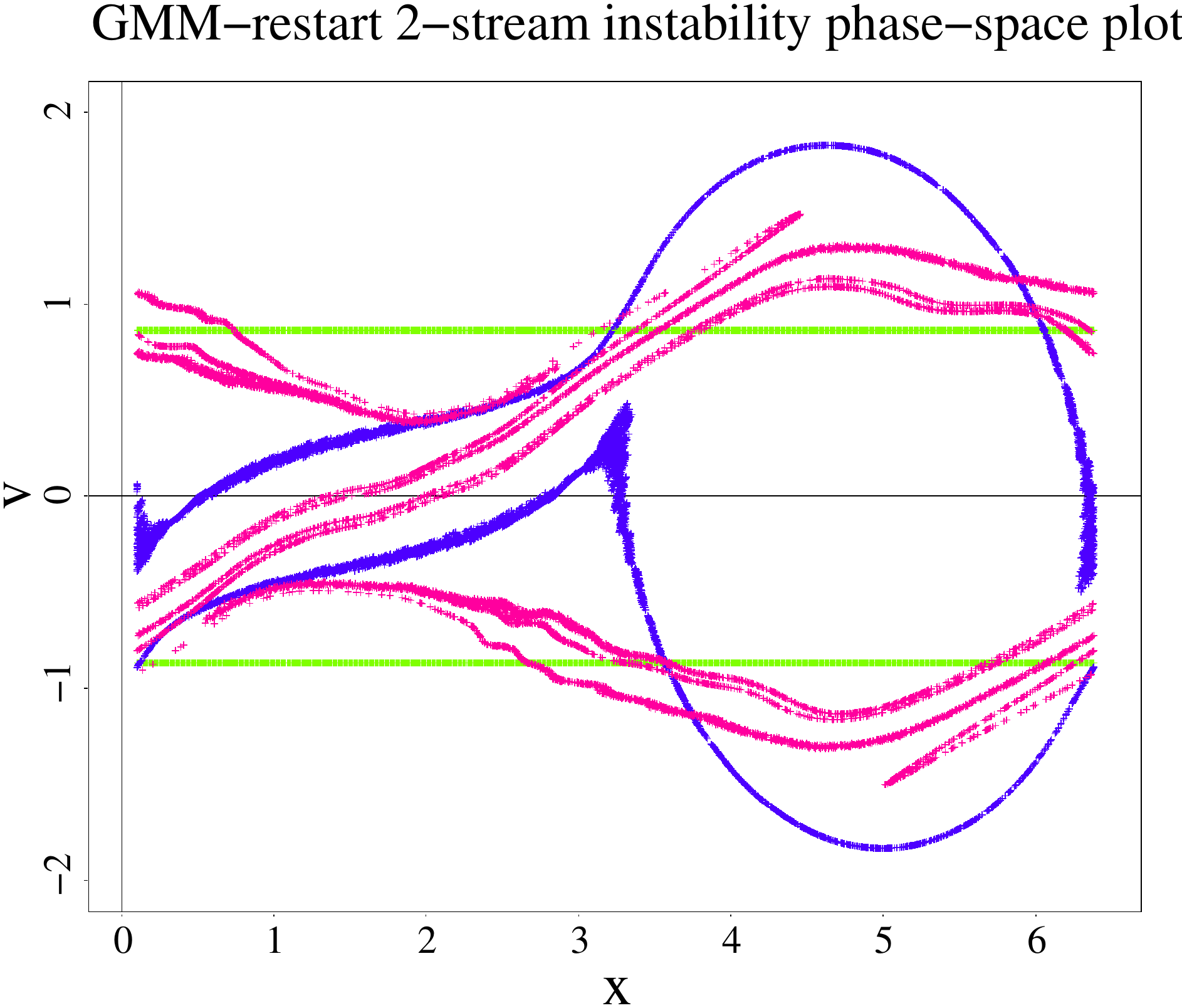}
\par\end{centering}
\caption{\label{fig:2-stream-phase-space}Two-stream instability: Phase-space
comparison at three different times (green: $t=0$, blue: $t=14.0$,
red: $t=19.4$) between the unrestarted case (left) and the GM-restarted
one (right).}
\end{figure}

A comparison between 1D-1V phase-space plots between unrestarted (left)
and GM-restarted (right) runs is shown in Fig. \ref{fig:2-stream-phase-space}.
It can be appreciated that the GM-restarted phase-space plot (right)
captures all phase-space features present in the unrestarted case
(left), except for a bit of beam-spread in the particles (which is
generated by the random uniform spatial initialization per cell in
the GM-restarted case).

\subsection{2D-3V Weibel electromagnetic instability}

Next we test with Weibel instability, which is a electromagnetic instability
in a plasma with anisotropic temperatures \citep{weibel1959spontaneously}.
Unless otherwise specified, simulations below are performed in a 2D
domain $10d_{e}\times10d_{e}$ (where $d_{e}$ is the electron skin
depth), with $16\times16$ cells, $\Delta t=1$ (in inverse plasma
frequency units), with doubly periodic boundary conditions. A temperature
anisotropy is set up for both electrons and ions with $v_{thx}=0.1$,
and $v_{thy,z}=0.3$ in speed-of-light units. The mass ratio is set
to be $m_{i}/m_{e}=1836$. We initialize the simulation with a $\delta$-function
perturbation in the particle velocities of $10^{-3}$, as described
in \citep{chen2015multi}. Note that the code employed has assumed
Darwin approximation \citep{chen2015multi}, which is non-relativistic,
and does not admit any light wave propagation in the system. 

\begin{figure}
\begin{centering}
\includegraphics[width=0.8\columnwidth]{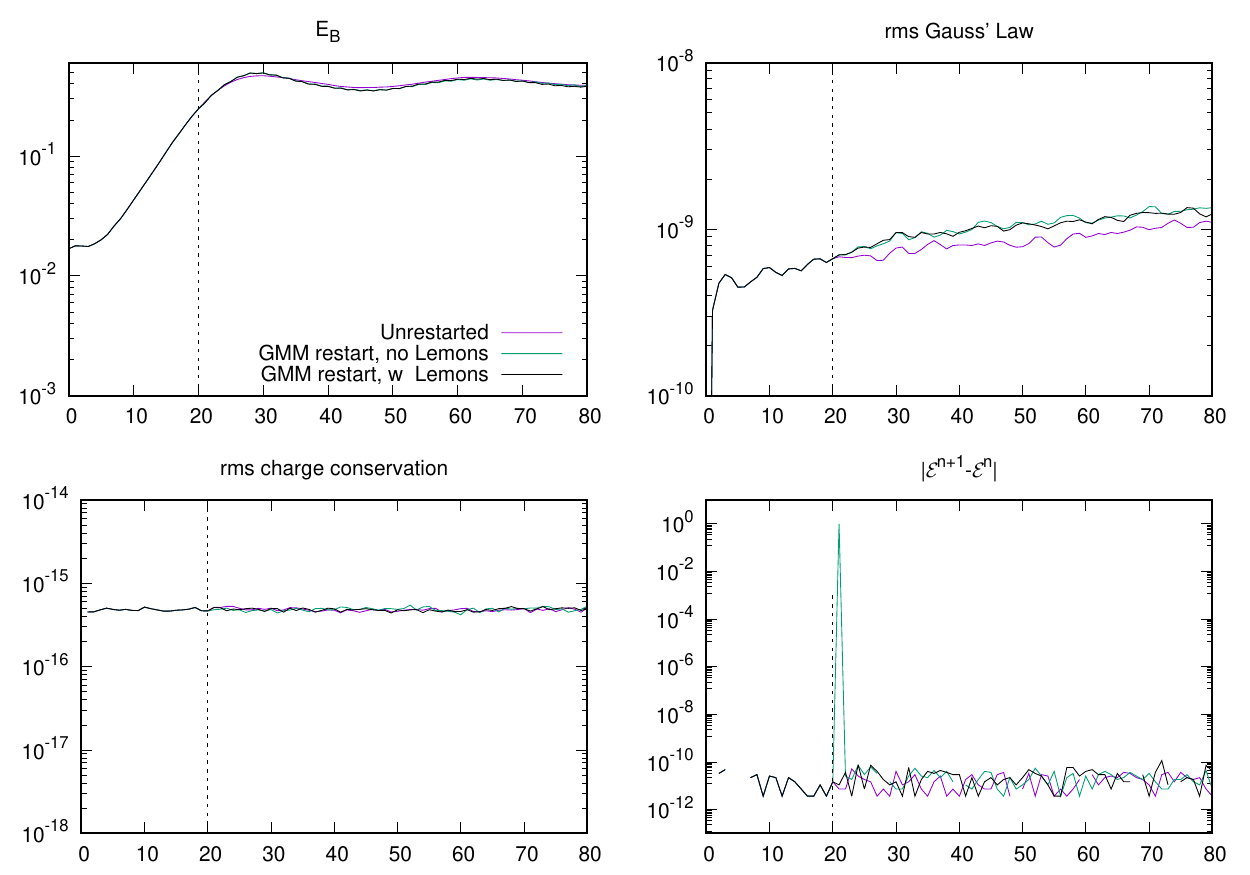}
\par\end{centering}
\caption{\label{fig:Weibel-instability-restarted-128ppc}2D Weibel instability
with $N_{p}=128$: Semi-log-scale time history of the magnetic field
energy $E_{B}$ (top-left), the rms of the Gauss' law residual over
the whole mesh (top-right), the rms of residual of the charge conservation
equation (bottom-left), and the change of total energy between subsequent
time steps (bottom-right). The simulations are obtained without restart,
and with GM restart at $t=20$ (in normalized units) with and without
Lemons moment matching. }
\end{figure}

\begin{figure}
\begin{centering}
\includegraphics[width=0.33\columnwidth,trim={0 4.5cm 6.5cm 0},clip]{./weibel2d-restart-trst=20-npc=128,128-qs=f.pdf}\includegraphics[width=0.33\columnwidth,trim={0 4.5cm 6.5cm 0},clip]{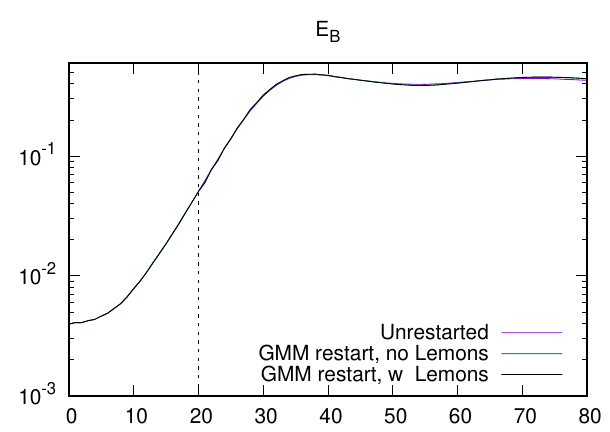}\includegraphics[width=0.33\columnwidth,trim={0 4.5cm 6.5cm 0},clip]{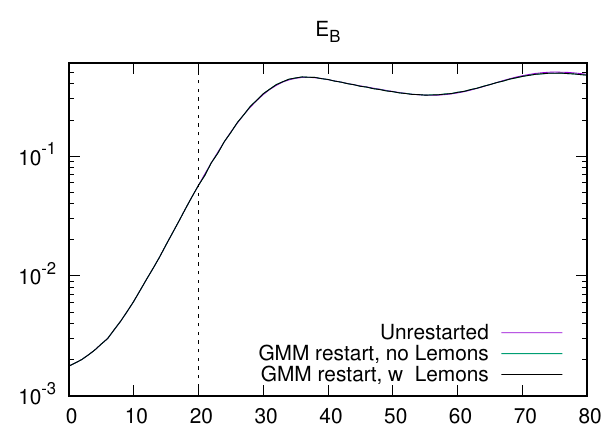}
\par\end{centering}
\caption{\label{fig:Weibel-instability-restarted-varying-Np}2D Weibel instability:
Semi-log-scale time history of the magnetic field energy with $N_{p}=128$
(left), 512 (center), and 1024 (right). The simulations are obtained
without restart, and with GM restart at $t=20$ (in normalized units)
with and without Lemons moment matching.}
\end{figure}

Figure \ref{fig:Weibel-instability-restarted-128ppc} shows similar
time histories as in Fig. \ref{fig:Two-stream-instability-restarted}
but with the magnetic energy $E_{B}=\sum_{i}B_{i}^{2}/2$, obtained
with $N_{p}=128$ particles per cell, and with and without restart
at $t=20$ in normalized time units (late in the linear phase). It
is apparent that conservation properties are preserved before and
after restart (except for energy without Lemons projection, as expected),
and that the GM restart quality is quite good, even with this relatively
small number of particles per cell. Increasing the number of particles
per cell improves the agreement, as it is shown in Fig. \ref{fig:Weibel-instability-restarted-varying-Np}.
The initial number of Gaussians per cell is 8, leading to an average
number of Gaussians per cell of 1.8, 2.1, 3.3 for 128, 512, and 1024
particles per cell, respectively, implying a compression ratio of
70, 240, and 310.

\textcolor{black}{}
\begin{figure}
\begin{centering}
\includegraphics[width=0.8\columnwidth]{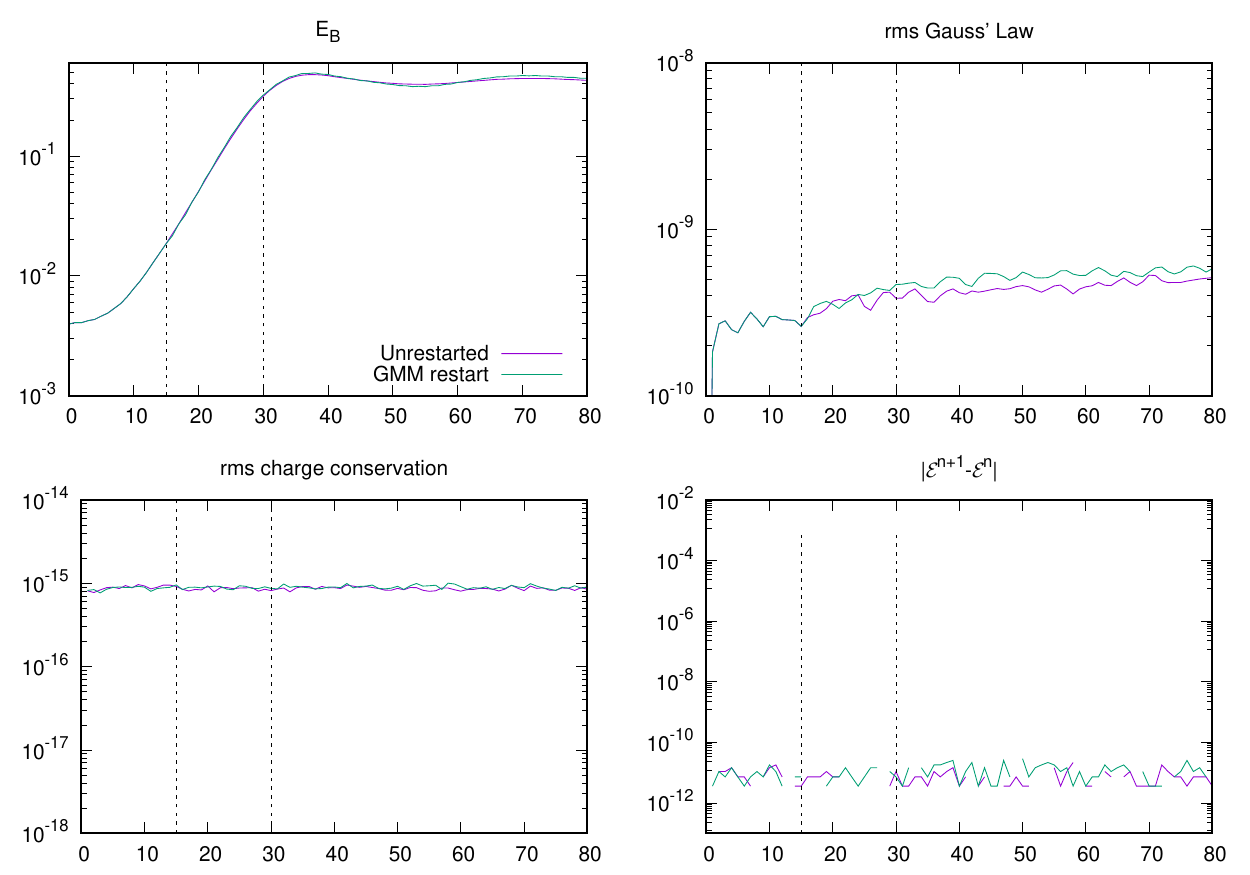}
\par\end{centering}
\textcolor{black}{\caption{\label{fig:variable-weight-dem}Time histories of the same quantities
as in Fig. \ref{fig:Weibel-instability-restarted-128ppc}, comparing
the twice-restarted 2D Weibel instability with 512 particles per cell
vs. the unrestarted result, demonstrating the ability of the method
to deal with particles of arbitrary weight.}
}

\end{figure}
\textcolor{black}{The ability of the approach to deal with particles
with different weights is shown in Fig. \ref{fig:variable-weight-dem},
which depicts similar time histories as before but now for the unrestarted
and twice-GM-restarted (at $t=15$ and $t=30$) 2D Weibel instability
with 512 particles per cell. The simulation begins with identical
particles, but their weights develop differences due to the density
mass-matrix solve at the first GM restart. The second GM restart is
therefore performed with non-identical particles. Agreement between
restarted and unrestarted time histories is very good throughout the
simulation, and demonstrates the ability of the method to deal with
particles with arbitrary weight.}

\subsection{Particle remapping using EM-GM for noise reduction (variance control)}

The central goal of machine-learning algorithms is not only to provide
a goodness-of-fit to the data, but also to be able to generalize.
The implication in the context of GM is that the estimated continuum
PDF may be able distinguishing between noise and signal, and, if so,
provide a measure of noise reduction (i.e., variance control), such
that the GM PDF, once resampled, may lead to an improved PIC solution
vs. the unrestarted one. The subject of noise control in PIC algorithms
has received significant attention recently \citep{wang11,myers20174th,faghihi2020moment},
but it has mostly been circumscribed to the remapping of the particle
PDF via interpolation  to a (semi-)structured phase-space mesh (i.e.,
bins), and subsequent resampling within bins. Some of these approaches
\citep{faghihi2020moment} explicitly embed arbitrary moment conservation
in their formulation, which is a desirable property. However, to our
knowledge, the use of Gaussian-mixture techniques for this purpose
remains unexplored.

Here, we provide anecdotal evidence that particle remapping using
the GM PDF reconstruction proposed here actually leads to an improvement
in PIC solution quality, suggesting that a thorough exploration of
this subject is worthwhile (and will be the subject of future work).
For our demonstration, we choose a Weibel instability in a 1D domain
of size $\pi$ (in $d_{e}$ units), with $\Delta t=1$ (in inverse
plasma frequency units), and periodic boundary conditions. The temperature
anisotropy and mass ratio is the same as in the previous Weibel example.
We initialize the simulation with a $\delta$-function perturbation
in the particle velocities of $10^{-2}$.

\begin{figure}
\begin{centering}
\includegraphics[width=1\columnwidth]{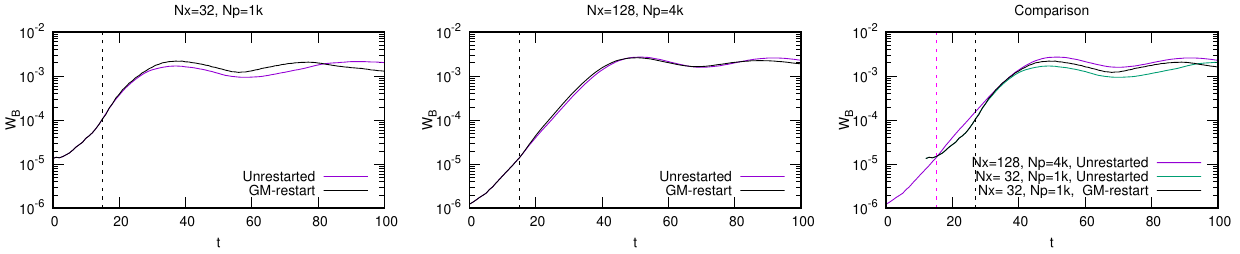}
\par\end{centering}
\caption{\label{fig:Weibel1d-convergence}1D Weibel instability comparison
with and without restart at $t=15$ for low resolution (left), high
resolution (center), and the comparison of the two (right). Note that,
for the right plot, a time shift ($t\leftarrow t+12$) has been applied
to the low-resolution histories to facilitate a meaningful assessment.}
\end{figure}

Figure \ref{fig:Weibel1d-convergence} shows a comparison of the magnetic-field
energy evolution between unrestarted and GM-restarted PIC simulations,
for a low-resolution case ($N_{p}=1000$, $N_{x}=32$, left) and a
high-resolution one ($N_{p}=4000$, $N_{x}=128$, center). Both are
restarted at $t=15$. For the low resolution case (Fig. \ref{fig:Weibel1d-convergence}-left),
one can appreciate a relatively big difference in the evolution of
the magnetic-field energy between the unrestarted and GM-restarted
simulations, especially when it enters the nonlinear stage. The magnetic
field energy is higher in the GM-restarted simulation, and there are
also some phase differences in the nonlinear oscillation. As we reduce
the grid size and increase the number of particles per cell, however,
the history of the magnetic-field energy agree much better between
unrestarted and GM-restarted solutions (Fig. \ref{fig:Weibel1d-convergence}-center),
indicating that the simulation is converging. More interestingly,
when one compares the low-resolution simulations with the high-resolution
ones (Fig. \ref{fig:Weibel1d-convergence}-right), it is apparent
that the GM-restarted low-resolution solution is much closer to the
high-resolution result than the unrestarted low-resolution one. It
follows that, for a given resolution, the GM-restarted simulation
is able to achieve a more accurate B-field nonlinear saturation energy
level than the unrestarted one, suggesting that the generalization
capability rooted in the unsupervised machine-learning algorithm is
in fact at play.

\subsection{On the acceleration of convergence of the EM-GM nonlinear algorithm}

\begin{figure}
\begin{centering}
\subfloat[$N_{p}=128$]{\begin{centering}
\includegraphics[width=0.33\columnwidth,trim=0.305in 0 0 0,clip]{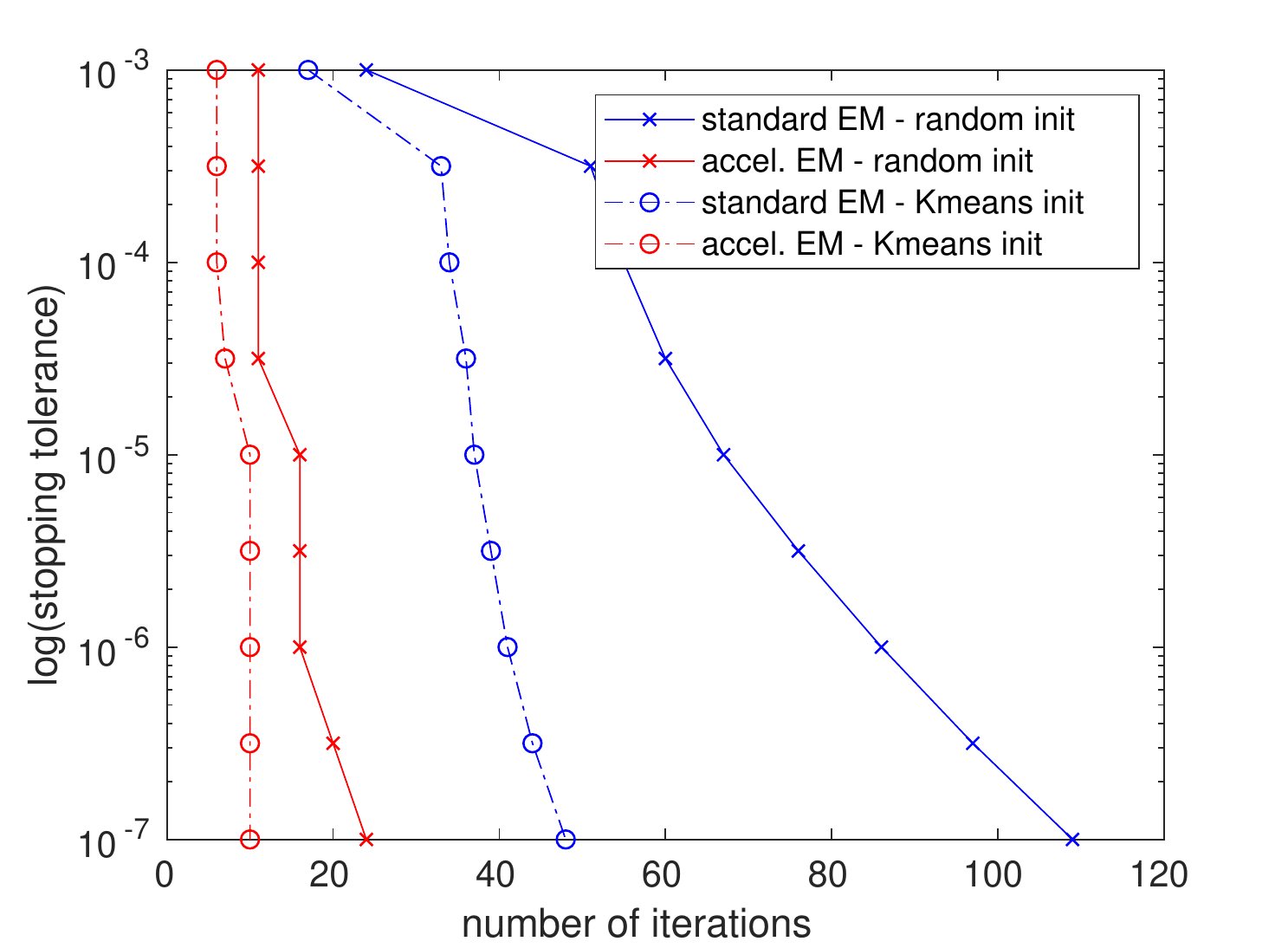}
\par\end{centering}
}\subfloat[$N_{p}=512$]{\begin{centering}
\includegraphics[width=0.33\columnwidth,trim=0.305in 0 0 0,clip]{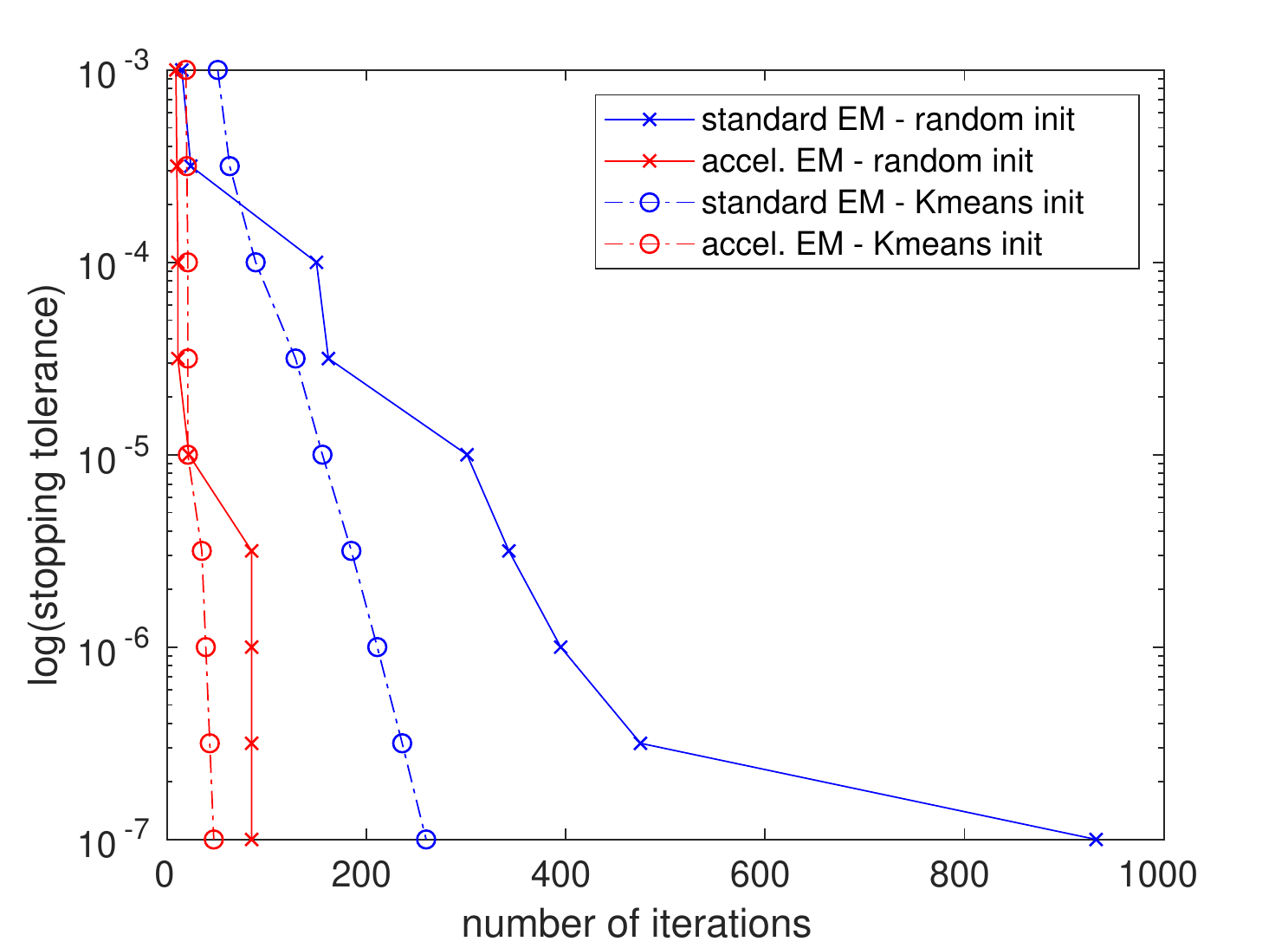}
\par\end{centering}
}\subfloat[$N_{p}=1024$]{\begin{centering}
\includegraphics[width=0.33\columnwidth,trim=0.305in 0 0 0,clip]{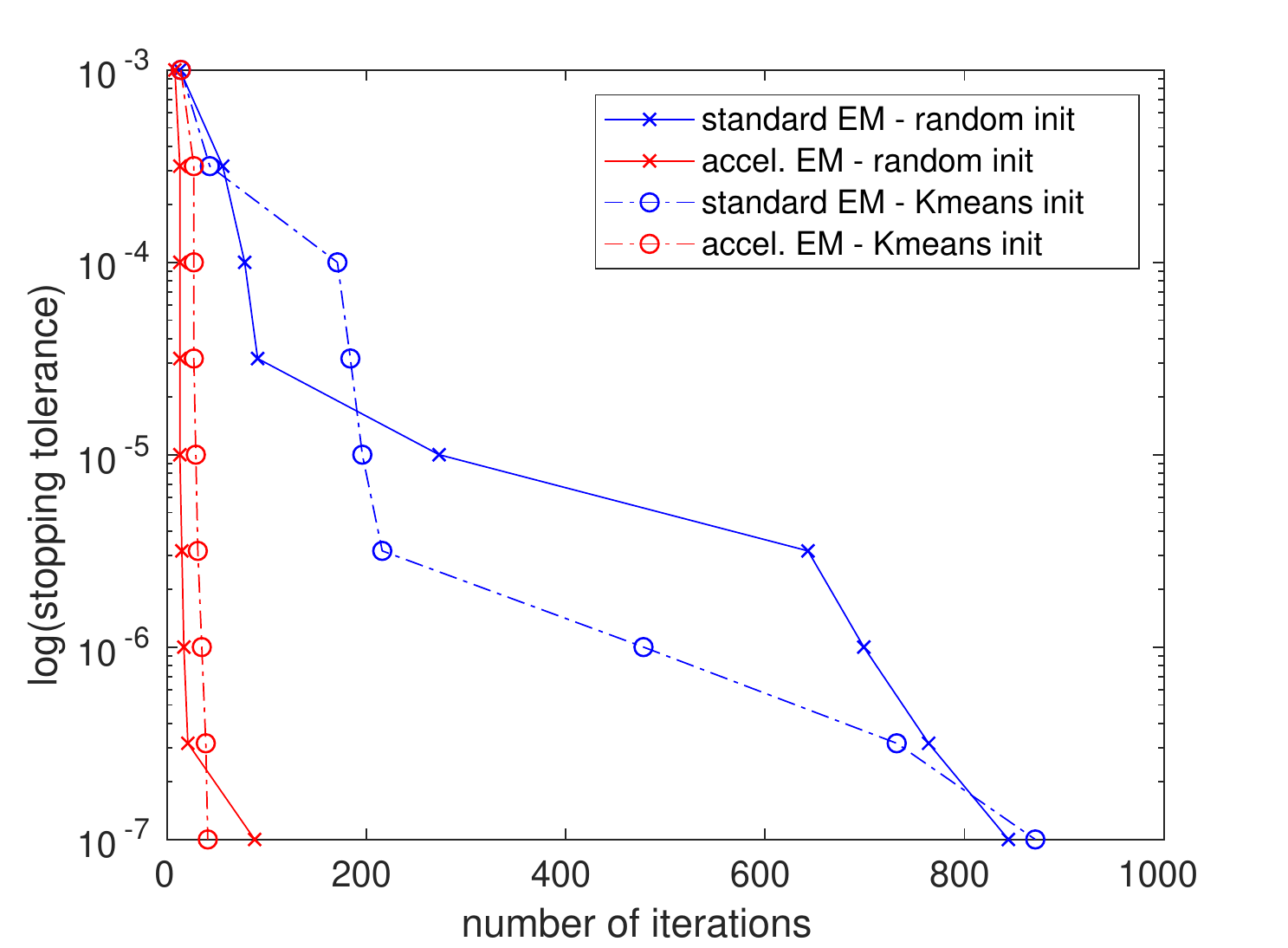}
\par\end{centering}
}
\par\end{centering}
\caption{\label{fig:Performance-of-EM}Performance of EM algorithm (relative
change of log-likelihood vs iteration) with and without AA and K-means
initialization at cell (1,3) of 2D Weibel problem at $t=20$ with
different particle resolutions.}
\end{figure}

As discussed previously, the EM-GM algorithm is guaranteed to converge
\citep{dempster1977maximum}, but performance can be slow. There has
been recent work trying to accelerate the convergence of the EM algorithm,
both by improving the initialization of the iteration (e.g., using
K-means \citep{plasse2013algorithm}), or by improving the Picard
iteration itself (e.g., by using Anderson Acceleration (AA) \citep{walker2011anderson,plasse2013algorithm}
or by advanced conjugate search direction algorithms \citep{he2012dynamic,xiang2020exact}).
In this study, we have implemented the K-means initialization and
the Anderson Acceleration algorithm in the \emph{standard} (non-adaptive)
EM-GM algorithm, and tested its impact using Weibel instability data.

A word is in order about our AA implementation for EM-GM, which to
our knowledge is new. We have included in the residual all degrees
of freedom for all Gaussians, namely, all weights $\omega_{k}$, means
$\boldsymbol{\mu}_{k}$, and second-moment matrices (i.e., $\rttensor{M}_{2,k}=\int d\mathbf{v}\,\mathbf{v}\mathbf{v}^{T}f_{k}$,
instead of covariance matrices, $\rttensor{\mathbf{\boldsymbol{\Sigma}}}_{k}=\int d\mathbf{v}(\mathbf{v}-\boldsymbol{\mu}_{k})(\mathbf{v}-\boldsymbol{\mu}_{k})^{T}f_{k}=\rttensor{M}_{2,k}-\boldsymbol{\mu}_{k}\boldsymbol{\mu}_{k}^{T}$).
The latter choice is motivated by the fact that $\rttensor{M}_{2,k}$
is linear in the mixture components (i.e., the second moment of a
linear combination of Gaussians is the linear combination of the second
moments of each individual Gaussian), whereas $\rttensor{\mathbf{\boldsymbol{\Sigma}}}_{k}$
is nonlinear, and therefore the former are better suited for acceleration
based on a linear combination of past residuals. It also has potential
advantages for the preservation of positivity of $\rttensor{\mathbf{\boldsymbol{\Sigma}}}_{k}$
(see below). It is important to note that, unlike EM, the standard
AA algorithm does not conserve moments, and does not guarantee that
covariance matrices remain positive definite. To fix conservation,
we apply a \emph{standard} EM step after the AA iteration (as was
done for the penalized EM algorithm for the same reason). To fix positivity,
we currently reset the AA iteration after an indefinite covariance
matrix is detected and revert back to a standard EM step. An alternate
approach (enabled by our choice to accelerate $\rttensor{M}_{2,k}$
instead of $\rttensor{\mathbf{\boldsymbol{\Sigma}}}_{k}$) would be
to guarantee that the Anderson mixing coefficients remain positive
(i.e., that the linear combination of residuals in AA remain convex).
Recent studies promote this as a viable globalization procedure for
AA \citep{chen2019convergence}, and this strategy will be explored
in future work. 

In Fig. \ref{fig:Performance-of-EM}, we report on the impact of these
strategies on the convergence rate of the for cell (3,1) of the 2D
Weibel test using 128, 512, and 1024 particles per cell. The plots
demonstrate that AA and K-means initialization (both independently
and combined) result in a significant speedup of the rate of convergence
of the non-adaptive EM-GM algorithm, which is more noticeable with
increasing number of particles per cell. It is apparent, however,
that AA is much more effective in accelerating the EM-GM convergence
than K-means. From these plots, the speedup between the unaccelerated
random initialization case to the AA accelerated K-means initialization
case is of more than an order of magnitude. These speedups are in
fact representative of performance in most cells for the Weibel test
problem. We are currently exploring ways of generalizing these strategies
for the \emph{adaptive} EM-GM algorithm, and these will be reported
in a future publication.

\section{Discussion and summary}

\label{sec:conclusions}

We have proposed a checkpoint-restart strategy for PIC algorithms
based on unsupervised machine-learning strategies using Gaussian Mixture
models. The Gaussian components are found adaptively using a penalized
Maximum Likelihood Estimate, solved by an Expectation-Maximization
procedure. For the numerical tests presented, the approach has demonstrated
significant compression potential (of several orders of magnitude)
without loss of physical fidelity (as demonstrated by actual restarted
PIC simulations). The latter is facilitated by the exact preservation
of charge, momentum, and energy in both compression and reconstruction
stages, and by the fact that GMM provides an optimal continuum reconstruction
of the PDF represented by the particles. Key to the fidelity of the
approach (particularly if many CRs are performed) is the use of a
mass-matrix procedure to match the density profile on the mesh exactly,
and a projection step to enforce conservation properties after particle
resampling. Our numerical experiments not only demonstrate that the
approach successfully restarts both electrostatic and electromagnetic
PIC simulations (with strict conservation of both charge and energy
exactly, and which therefore represent a stringent test of the method),
but also suggest that a periodic GM particle remap may in fact improve
the quality of PIC solutions. This point, which is anecdotal in this
study, suggests the possibility of machine-learning variance reduction
in particle methods, and will be investigated further in future work.
Finally, we have proposed a simple implementation strategy for Anderson
Acceleration in the \emph{non-adaptive} EM-GM algorithm that results
in convergence speedups of more than an order of magnitude, while
strictly preserving the positivity of the covariance matrices of the
mixture. Beyond CR and particle remapping for enhanced solution quality,
we note that the approach outlined in this study enables straightforwardly
particle redistribution over the computational domain, to facilitate
various performance goals such as load balancing and particle-number
control. This will also be the subject of future studies.

\section*{Acknowledgements }

This work was supported by the U.S. Department of Energy, Office of
Science, Office of Applied Scientific Computing Research (ASCR), both
by the EXPRESS (2016-17) and SciDAC (2018-20) programs. This research
used resources provided by the Los Alamos National Laboratory Institutional
Computing Program, and was performed under the auspices of the National
Nuclear Security Administration of the U.S. Department of Energy at
Los Alamos National Laboratory, managed by Triad National Security,
LLC under contract 89233218CNA000001.

\appendix

\section{Derivation of Penalized Likelihood Function\label{app:PLF}}

\textcolor{black}{We begin by making use of the exact decomposition
\citep{neal1998view}:
\begin{equation}
\mathrm{ln}p(\mathbf{X}|K)=L(q)+KL(q||p),
\end{equation}
where 
\begin{align}
L(q) & =\int q(\bm{\theta})\mathrm{ln}\frac{p(\mathbf{X}|\bm{\theta},K)p(\bm{\theta}|K)}{q(\bm{\theta})}d\bm{\theta},\label{eq:lower bound}\\
KL(q||p) & =\int q(\bm{\theta})\mathrm{ln}\frac{q(\bm{\theta})}{p(\bm{\theta}|\mathbf{X},K)}d\bm{\theta},\label{eq:KL}
\end{align}
Note that the decomposition holds for an arbitrary positive-definite
distribution $q(\bm{\theta})$. Since the Kullback-Leibler divergence
$KL(q||p)$ is always greater or equal to zero \citep{mackay2003information},
$L(q)$ is a lower bound of the log-marginal likelihood. In fact,
maximizing $L(q)$ is equivalent to maximizing $\mathrm{ln}p(\mathbf{X}|K)$
\citep{neal1998view}. Various forms of $q(\bm{\theta})$ can be adopted.
For instance, variational Bayesian methods assume that $q(\bm{\theta})$
factorizes over subsets $\{\bm{\theta}_{i}\}$, i.e., $q(\bm{\theta})=\Pi_{i}q_{i}(\bm{\theta}_{i})$
\citep{corduneanu2001variational}. Here, $q(\bm{\theta})$ is assumed
to be a uniform distribution in a small interval $(\mathbf{a},\mathbf{b})$
around a point of $\boldsymbol{\theta}$, i.e., $q(\bm{\theta})=1/\Delta$,
with $\Delta\equiv\int_{\mathbf{a}}^{\mathbf{b}}d\bm{\theta}$ the
volume of a $d$-dimensional hypercube in the space of parameter $\boldsymbol{\theta}$.
Equation \ref{eq:lower bound} then becomes 
\begin{equation}
L(\Delta)=\frac{1}{\Delta}\int_{\mathbf{a}}^{\mathbf{b}}\mathrm{ln}\left[p(\bm{\theta}|K)\Delta\right]d\bm{\theta}+\frac{1}{\Delta}\int_{\mathbf{a}}^{\mathbf{b}}\mathrm{ln}\left[p(\mathbf{X}|\bm{\theta},K)\right]d\bm{\theta}.\label{eq:lower bound w/ uniform q}
\end{equation}
In the context of information theory, the maximum of Eq. \ref{eq:lower bound w/ uniform q}
is equivalent to the shortest message length that the data can communicate
\citep{mackay2003information}. The idea is that the model with the
minimum message length (thus so-called MML) should be preferred. The
message length, defined as $l(x)=-\mathrm{ln}P(x)$, where $P(x)$
is the probability of an event $x$, is a measure of the information
content of the event $x$ \citep{mackay2003information}. For a continuous
PDF $p(x)$, $l(x)=-\mathrm{ln}[p(x)dx]$, with $dx$ a small interval
around $x$. It is clear from this perspective that the first term
on the right-hand-side (rhs) of Eq. \ref{eq:lower bound w/ uniform q}
corresponds to the message length of $\bm{\theta}$, and $\Delta$
denotes a discretization of $\bm{\theta}$ (which may be thought as
the finite precision of $\bm{\theta}$). The finite precision of $\bm{\theta}$
has a major effect on ``communicating'' the message length of the
data. Expectation is taken with respect to the assumed uniform distribution
$q(\bm{\theta})$ over a small interval $(\mathbf{a},\mathbf{b})$,
and an optimum $\Delta$ can be found by maximizing Eq. \ref{eq:lower bound w/ uniform q}.}

We next rewrite the log-marginal likelihood function (Eq. \ref{eq:lower bound w/ uniform q})
as:
\begin{equation}
L(\Delta_{\xi})=\frac{1}{\Delta_{\xi}}\int_{\boldsymbol{\alpha}}^{\mathbf{\boldsymbol{\beta}}}\mathrm{ln}\left[p(\bm{\xi}|K,\boldsymbol{\omega})p(\boldsymbol{\omega})\Delta_{\xi}\right]d\bm{\xi}+\frac{1}{\Delta_{\xi}}\int_{\mathbf{\boldsymbol{\alpha}}}^{\mathbf{\boldsymbol{\beta}}}\mathrm{ln}\left[p(\mathbf{X}|\bm{\xi},K,\boldsymbol{\omega})\right]d\bm{\xi}.\label{eq:lower bound w/ uniform q-1}
\end{equation}
where we have made a variable transformation $\bm{\theta}=\Lambda^{-1/2}U^{T}\bm{\xi}$
with Jacobian $J=|\partial\boldsymbol{\theta}/\partial\boldsymbol{\xi}|$.
Here $U$ is a $d\times d$ orthogonal matrix with columns given by
eigenvectors and $\Lambda$ is a $d\times d$ diagonal matrix with
elements of eigenvalues of the observed Fisher information matrix,
$I_{p}=-\left.\frac{\partial^{2}\mathrm{ln}p(\mathbf{X}|\bm{\theta},K,\boldsymbol{\omega})}{\partial\bm{\theta}^{2}}\right|_{\tilde{\bm{\theta}}}$.
Here we have assumed that the Hessian matrix is a negative semidefinite
(e.g., when $\mathrm{ln}p$ is concave), so that we can write $I_{p}=U\Lambda U^{T}$,
where $U$ is an orthogonal matrix. It follows that $J=|I_{p}|^{-\frac{1}{2}}$.
Using the chain rule \citep{schervish2012theory} we find that $I_{p}(\boldsymbol{\xi})=(\partial\boldsymbol{\theta}/\partial\boldsymbol{\xi}){}^{T}I_{p}(\partial\boldsymbol{\theta}/\partial\boldsymbol{\xi})=\mathbb{1}$.
A truncated Taylor expansion with respect to the center (denoted as
$\tilde{\bm{\xi}}$) of $\Delta_{\bm{\xi}}$ is typically employed
to approximate the log-likelihood $\mathrm{ln}p(\mathbf{X}|\bm{\xi},K,\omega)$:
\begin{equation}
\mathrm{ln}p(\mathbf{X}|\bm{\xi},K,\omega)\simeq\mathrm{ln}p(\mathbf{X}|\tilde{\bm{\xi}},K,\omega)+(\bm{\xi}-\tilde{\bm{\xi}})\cdot\left.\frac{\partial\mathrm{ln}p}{\partial\bm{\xi}}\right|_{\tilde{\bm{\xi}}}+\frac{1}{2}(\bm{\xi}-\tilde{\bm{\xi}})^{T}\cdot(\bm{\xi}-\tilde{\bm{\xi}}).\label{eq:Taylor expansion}
\end{equation}
Substituting Eq. \ref{eq:Taylor expansion} into Eq. \ref{eq:lower bound w/ uniform q-1}
results
\begin{equation}
L(\tilde{\bm{\xi}},\Delta_{\bm{\xi}})=\mathrm{ln}\left[p(\mathbf{X}|\tilde{\bm{\xi}},K,\omega)p(\tilde{\bm{\xi}}|K,\omega)p(\omega)\Delta_{\bm{\xi}}\right]-\frac{d}{24}\Delta_{\bm{\xi}}^{2/d},\label{eq:lower bound w/ uniform xi}
\end{equation}
where we have used $\frac{1}{\Delta_{\bm{\xi}}}\int_{\boldsymbol{\alpha}}^{\boldsymbol{\beta}}(\bm{\xi}-\tilde{\bm{\xi}})d\bm{\xi}=0$,
and $\frac{1}{\Delta_{\bm{\xi}}}\int_{\boldsymbol{\alpha}}^{\boldsymbol{\beta}}(\bm{\xi}-\tilde{\bm{\xi}})^{T}\cdot(\bm{\xi}-\tilde{\bm{\xi}})d\bm{\xi}=\frac{d}{12}\Delta_{\bm{\xi}}^{2/d}$,
both integrated over the volume $\Delta_{\bm{\xi}}$ (a $d$-dimensional
hypercube), and assuming that all the other terms are constant within
$\Delta_{\bm{\xi}}$. By setting $\frac{\partial L}{\partial\Delta_{\bm{\xi}}}=0$,
$\Delta_{\bm{\xi}}=(12)^{d/2}$ is found to maximize Eq. \ref{eq:lower bound w/ uniform xi}.
Substituting $\Delta_{\bm{\xi}}=(12)^{d/2}$ into Eq. \ref{eq:lower bound w/ uniform xi}
yields: 
\begin{equation}
L(\tilde{\bm{\theta}})=\mathrm{ln}\left[p(\mathbf{X}|\tilde{\bm{\theta}},K,\omega)p(\tilde{\bm{\theta}}|K,\omega)p(\omega)\right]-\frac{1}{2}\mathrm{ln}|I_{p}|-\frac{d}{2}(1-\mathrm{ln}12),\label{eq:minus MML criterion}
\end{equation}
where we have also used $p(\mathbf{X}|\tilde{\bm{\theta}},K)p(\tilde{\bm{\theta}}|K)J=p(\mathbf{X}|\tilde{\bm{\xi}},K)p(\tilde{\bm{\xi}}|K)$
due to the variable transformation \citep{casella2002statistical}.
The negative of Equation \ref{eq:minus MML criterion} is the so-called
MML criterion \citep{lanterman2001schwarz}:
\begin{equation}
Message\:Length=-\mathrm{ln}\left[p(\mathbf{X}|\tilde{\bm{\theta}},K,\omega)p(\tilde{\bm{\theta}}|K,\omega)p(\omega)\right]+\frac{1}{2}\mathrm{ln}|I_{p}|+\frac{d}{2}(1-\mathrm{ln}12).\label{eq:MML criterion}
\end{equation}
Note that $\Delta_{\bm{\xi}}$ related terms group into the last term,
which is in general not that important (see below).

We must further simplify Eq. \ref{eq:minus MML criterion} because
of the difficulties in selecting prior distributions \citep{kass1996selection}
and calculating the Fisher information. We start with noting that
$I_{p}=-\left.\frac{\partial^{2}\mathrm{\sum_{i=1}^{N}ln}p(\mathbf{x}_{i}|\bm{\theta},K)}{\partial\bm{\theta}^{2}}\right|_{\tilde{\bm{\theta}}}=-\sum_{i=1}^{N}\left.\frac{\partial^{2}\mathrm{ln}p(\mathbf{x}_{i}|\bm{\theta},K)}{\partial\bm{\theta}^{2}}\right|_{\tilde{\bm{\theta}}}\simeq N\mathcal{I}$
where $\mathcal{I}=-E_{\boldsymbol{\theta}}(\partial^{2}\mathrm{ln}p(\mathbf{x}|\bm{\theta},K)/\partial\theta^{2})$\textcolor{black}{{}
is the Fisher information matrix (FIM)}, and the expectation is taken
with the mixture PDF \citep{mclachlan2004finite}. To proceed, only
an upper bound of $|\mathcal{I}|$, the complete FIM $\mathcal{I}_{c}$
\citep{titterington1985statistical}, is considered \citep{raim2017approximation},
\begin{equation}
\mathcal{I}_{c}=\mathrm{Blockdiag}(\omega_{1}I_{1},...,\omega_{K}I_{K},I_{\omega}),
\end{equation}
where $I_{k}$ is a $T\times T$ FIM, with $T=\frac{1}{2}D(D+3)$
and recall that $D$ is the dimension of $\bm{\mu}$. Note that $d\times d$
is the dimension of $I$, where $d=KT+K-1$ is the total number of
parameters. The minus one is due to the constraint that $\sum_{i}\omega_{i}=1$.
The second term on the rhs of Eq. \ref{eq:minus MML criterion} may
be written as $-\frac{1}{2}\mathrm{ln}(N^{d}|\mathcal{I}_{c}|)=-\frac{1}{2}\mathrm{ln}[N^{d}\prod_{i=1}^{K}(\omega_{i}^{T}|I_{i}|)|I_{\omega}|]$.
With the above approximations, Eq. \ref{eq:minus MML criterion} yields:
\begin{equation}
L(\tilde{\bm{\theta}})=\mathrm{ln}\left(\frac{p(\tilde{\bm{\theta}}|K)}{\sqrt{|I_{\omega}|}\prod_{i=1}^{K}\sqrt{|I_{i}|}}\right)+\mathrm{ln}p(\mathbf{X}|\tilde{\bm{\theta}},K)-\frac{d}{2}\mathrm{ln}N-\frac{T}{2}\sum_{i=1}^{K}\mathrm{ln}\omega_{i}-\frac{d}{2}(1-\mathrm{ln}12).
\end{equation}
If we choose independent priors, i.e., $p(\tilde{\bm{\theta}}|K)=p(\bm{\omega})\prod_{i=1}^{K}p(\bm{\mu}_{i},\bm{\Sigma}_{i})$,
and adopt Jeffreys' prior for $\bm{\omega}$ and for each $(\bm{\mu}_{i},\bm{\Sigma}_{i})$
\citep{bernardo1988bayesian,figueiredo2000unsupervised}, we obtain
\begin{equation}
L(\bm{\omega},\bm{\mu},\rttensor{\mathbf{\boldsymbol{\Sigma}}})=\mathrm{ln}\left(p(\mathbf{X}|\bm{\omega},\bm{\mu},\rttensor{\mathbf{\boldsymbol{\Sigma}}},K)\right)-\frac{d}{2}\mathrm{ln}N-\frac{T}{2}\sum_{i=1}^{K}\mathrm{ln}(\omega_{i}),\label{eq:Simple penalized likelihood-1}
\end{equation}
after dropping some constants and $\sim O(d)$ terms (more specifically
the condition for dropping the last term of Eq. \ref{eq:minus MML criterion}
is $N\gg1$ which is typically the case). We have arrived at a simple
penalized likelihood function, Eq. \ref{eq:Simple penalized likelihood-1}.
It is worth noting that the MML estimator of maximizing the likelihood
Eq. \ref{eq:Simple penalized likelihood-1} is invariant under variable
transformation, or re-parameterization of $(\bm{\mu},\rttensor{\mathbf{\boldsymbol{\Sigma}}})$
and $\boldsymbol{\omega}$. This is due to the invariance property
of maximum likelihood estimators to arbitrary transformations of the
parameters of likelihood function \citep{casella2002statistical}.

\pagebreak\bibliographystyle{ieeetr}
\bibliography{GMM}

\end{document}